\documentclass[12pt,nofootinbib, longbibliography]{revtex4-1}
\usepackage{amsmath,amssymb}
\usepackage{slashed,verbatim,graphicx}
\usepackage{graphicx}
\usepackage{ytableau}
\usepackage{pstricks}

\usepackage{color}

\newcommand\vev[1]{\langle #1\rangle}
\newcommand\ket[1]{| #1\rangle}
\newcommand\bra[1]{\langle #1|}
\newcommand\braket[2]{\langle #1|#2\rangle}
\newcommand\BR{\mathbb{R}}

\newcommand{\BC}{\mathbb {C}}

\def\Tr{\textrm{Tr}}

\newcommand{\CO}{  {\cal O}  }

\begin{document}
\setlength{\unitlength}{1mm}
\title{Code subspaces for LLM geometries}

\author{David Berenstein, Alexandra Miller}
\affiliation { Department of Physics, University of California at Santa Barbara, CA 93106}

\begin{abstract} 
We consider effective field theory around  classical background geometries with a gauge theory dual, in the class  of LLM geometries. These  are dual to half-BPS states of $\cal{N}=$ 4 SYM. We find that the language of code subspaces is natural for discussing the set of nearby states, which are built by acting with effective fields on these backgrounds. This work extends our previous work by going beyond the strict infinite $N$ limit. We further discuss how one can extract the topology of the state beyond $N\rightarrow\infty$ and find that uncertainty and entanglement entropy calculations still provide a useful tool to do so. Finally, we discuss obstructions to writing down a globally defined metric operator. We find that the answer depends on the choice of reference state that  one starts with. Therefore there is ambiguity in trying to write an operator that describes the metric globally.

 \end{abstract}

\maketitle

\section{Introduction }
\label{S:Introduction}

The existence of gauge/gravity dualities \cite{Maldacena:1997re, Witten:1998qj,Gubser:1998bc} is remarkable and with each passing day we discover a new piece to their puzzle. Part of why these theories are so mysterious is because they often have non-intuitive and surprising properties that seem to lead to paradoxes. One of these was recently resolved by Almheiri, Dong, and Harlow \cite{Almheiri:2014lwa}. The puzzle they addressed was that a local field at a point in the bulk should have vanishing commutators with fields that are spatially separated from them, including the boundary. A point in the center of global AdS would be spatially separated from the boundary at $t=0$ (they belong to the same Cauchy slice) and would therefore have to commute with all local operator insertions on the boundary. Such a field should act trivially on the Hilbert space of states, and yet, be encoded as a non-trivial operator on the boundary.  
That is,  bulk information seemed to be non-localized in the boundary theory in an unexpected way. Their resolution was that though this might seem strange, it is not an entirely new phenomenon and in fact has a very nice interpretation in terms of quantum information theory.  The idea is that the holographic correspondence acts as a quantum error correcting code. The commutation properties that are required are true in a subspace of the Hilbert space called the code subspace. It is this restriction to the code subspace that makes it possible to have the vanishing commutators in an effective sense, rather than as a statement on the full Hilbert space of states, where such properties are forbidden by quantum field theory theorems.
Since their original work, there has been much progress in reinterpreting gauge/gravity dualities as holographic codes (for instance \cite{Mintun:2015qda, Pastawski:2015qua, Freivogel:2016zsb}). In this paper, we will push the idea further. We find that the language of code subspaces is a natural home for effective field theory and further, this can be seen explicitly within the framework of the LLM geometries \cite{Lin:2004nb}.

In standard quantum information theory, if one wants to encode a message, one utilizes a Hilbert space of states larger  than is necessary. One then constrains allowed messages to a particular subspace, the code subspace. For instance, one might use a few qubits to send a one qubit message, this provides a larger Hilbert space to work with and allows the messenger to choose whatever subspace they like to work within. In \cite{Almheiri:2014lwa}, the authors defined a set of code subspaces in AdS/CFT to be those formed as the linear span of

\begin{equation}
\ket\Omega, \, \phi_i(x) \ket\Omega, \,\phi_i(x_i) \phi_j(x_2) \ket\Omega, \dots \label{eq:subspace}
\end{equation}
where the $\phi_i(x)$ make up some finite set of local bulk operators, which can be realized in the CFT with the Hamilton-Kabat-Lifschitz-Lowe  reconstruction of bulk operators \cite{Hamilton:2005ju}. Almheiry et al. take $\ket\Omega$ to be the ground state of the system (though they say one could also consider other semiclassical background states, as we will do explicitly). The quantum error correcting properties of the gauge/gravity duality code make it possible to have a realization of the fields $\phi$ in the code subspace that commute with operators on the boundary, as long as one restricts the evaluation to states that belong to the code subspace.

In this work, we start by discussing effective field theory around a given classical background. We consider the Hilbert space accessible to an experimenter, which can be built by acting on the background state with some set of effective fields, in a way similar to equation \eqref{eq:subspace}. 
That is, we want to take a construction similar to the code subspace of vacuum AdS within the confines of effective field theory and show that this has a lot of desirable properties for addressing more general questions of quantum gravity.  
Because an experimenter will not have access to infinite energy, they cannot act with all fields in the theory, but rather they are constrained by some cutoff (both in momentum and occupation number). It will turn out that the details of what cutoff is appropriate for each background will depend on the particular background under consideration. That is, the formulation of effective field theory is state dependent in relation to the reference state that we choose to expand from. We observe that the space built in this way exactly matches the structure of the code subspace defined in \cite{Almheiri:2014lwa} and we expand on this fact. This is also very similar to how the Hilbert space of nearby states is built around black hole states in the work of Papadodimas and Raju \cite{Papadodimas:2012aq, Papadodimas:2013jku}, by starting with a reference state. We discuss this philosophy in section \ref{sec:eff}.

In the rest of the paper, we deal specifically with the example of the LLM Geometries, which are dual to the half-BPS states of $\cal{N}=$ 4 SYM. This is a very useful  setup to work with as it is a place where we understand both the geometric description \cite{Lin:2004nb} and the field theory Hilbert space well \cite{Corley:2001zk,Berenstein:2004kk}. In fact, most of the exact  computations turn out to be  combinatorial in nature \cite{Berenstein:2017abm} (see also \cite{Lin:2017dnz}). In section \ref{sec:YT}, we build some necessary technology to do computations. The basic input is that there are two convenient bases for describing the set of half-BPS states: one that can be classified by Young diagrams and one built by taking traces of powers of a matrix $Z$. In this section, we provide the details of these two bases, describing how to go between them and computing their inner products. These will be the tools we need for the remainder of the work, and generalize some constructions that were carried out in our previous work \cite{Berenstein:2017abm} for finite $N$. 

These states are dual to the set of half-BPS states in type IIB supergravity: the LLM geometries. Each geometry in this set can be classified by a black and white coloring of the plane. We consider the set of concentric ring configurations, because they are dual to states that are simple to describe in terms of Young diagrams \cite{Lin:2004nb,Mosaffa:2006qk}. Although in principle other such geometries could be analyzed, the control of the states in the field theory dual is  poor and relies on approximations. With the concentric configurations, we can make exact statements in the Hilbert space of states.
 These states dual to concentric rings will be the background states upon which we build our code subspaces. The nearby states that make up the subspace are built by acting with effective gravity field perturbations on each edge of the rings, which causes them to deform. With a bit of work, we are able to write down these fields explicitly so that we can build the states as in \eqref{eq:subspace} in a way that is suitable for our purposes. This is the content of section \ref{sec:geomcode}.

By only considering the geometric description, the aforementioned cutoffs are not immediately clear and although in principle one should  be able to derive them, it takes a considerable amount of effort. So, instead, in section \ref{sec:codeyoung}, we go back to the representation of the states in terms of Young diagrams, where things become clearer. The well behaved concentric ring configurations correspond to diagrams with few corners. The effective fields are build out of modes that act only in a particular corner of the diagram and one can reproduce directly the supergravity analysis entirely with combinatorial techniques. These techniques have been developed in various papers for different settings (see \cite{Kristjansen:2002bb,Koch:2008ah, Berenstein:2013md, Berenstein:2017abm,Lin:2017dnz} and references therein).

In section \ref{sec:cutoff}, we undertake the problem of understanding the cutoff. The cutoff is provided by constraining the excitations so that they do not simultaneously affect multiple corners of the diagram and by requiring that they are sufficiently {\em planar}. This corresponds to each field only acting on a single edge of the concentric rings and having small energy, although the energy of an individual quantum can be much larger than the Planck scale\footnote{The energy of an individual quantum can scale like $N^{1/2}$, rather than $N^{1/4}$}. We can write these excitations in terms of modes that act in a particular corner of the Young diagram and from these build a Fock space representation.

In section  \ref{sec:uncent} we expand on some of our previous work \cite{Berenstein:2017abm, Berenstein:2016pcx}, where we  compute the topology of the states within a given code subspace. Previously, we were in the strict $N\rightarrow\infty$ limit, but here we  go beyond that, taking $N$ to be large, but finite. As before, we find that we can extract the topological information from entanglement and uncertainty calculations, though it requires more work: a number can not be guessed any longer form a single mode, but it requires many modes instead.
Here we also find a close connection with the recent work of Balasubramanian, et. al. \cite{Balasubramanian:2017hgy}, who showed the existence of entanglement shadows in the LLM geometries. We find that similarly, the extrapolate dictionary seems to stop at the outermost anti-edge of the concentric ring diagrams. This is the second edge starting from the outside going inward in the radial direction of the LLM plane and it is the same place where the entanglement shadow begins.

Finally, in section \ref{sec:obs} we consider the overlap that can occur between different code subspaces. We look specifically at an example where you start with two different background states and add particular excitations to each, which results in having prepared two identical states from the viewpoint of Young diagrams, but whose construction indicates that they should be assigned different metric operators. We discuss the ambiguities that arise because of this fact, which in particular obscures one's ability to write down a globally well defined metric operator.

\section{Code subspaces and effective field theory}\label{sec:eff}

In this section, we will discuss doing effective field theory around some classical background. We will consider the constraints put on an experimenter in this set-up and will show how what we end up with matches previous definitions of code subspaces.

Let us start by assuming we are given a quantum state  $\ket B$ that is dual to a classical background for a field or gravitational theory. Eventually we will work in a field theory with a gravitational dual, using the gauge/gravity duality. Here, the state $\ket B$ will correspond to a classical background in the bulk, rather than the boundary theory.  Though $\ket B$ is a classical background, we need to be careful, because in the quantum theory the quantum fluctuations can never be zero. Instead, we should think of $\ket B$ as a coherent state, where (effective) quantum fields have minimal uncertainty relative to the background.
 We also want to be careful because we will often have a cutoff to account for. For instance, if $\ket B$ is a ground state for a gapped system, $\ket 0$, the cutoff might be in the energy available to us. This will restrict us in two different ways. First, it will require that the only modes that can be excited are long wavelength fluctuations (of small enough energy) and further, we will be restricted in the occupation number of any one such
 mode, so that the energy cutoff also imposes an amplitude cutoff for any one mode.
 Generally, this could be configuration dependent if for example, the gap for some additional excitation depends on a vacuum expectation value. This is common in supersymmetric field theories when we have a moduli space of vacua. 

In analyzing this system, we might want to understand what a Hilbert space of {\em nearby} states to the background $\ket B$ would look like. The state $\ket B$ belongs to a Hilbert space of states ${\cal H}$ that defines the full quantum theory. It is tempting to consider the set of states  $\ket \psi \in {\cal H}$ such that they differ from $\ket B$ by a small amount,  $\epsilon$ inside the Hilbert space $|\ket \psi -\ket B|< \epsilon$. There are many problems with this prescription, and we will enumerate a few of them in what follows. First, the set of states $\ket \psi$ is not a linear subspace of ${\cal H}$: we cannot do quantum mechanics restricted to the nearby states. Secondly, the set of such states $\ket \psi$ makes no mention of the cutoff nor to the effective fields. 

We want to define the set of {\em nearby} states to be those that can be generated from $\ket B$ by the action of the effective fields, and so that it is also a linear space. That is, we want the set of nearby states to be a Hilbert space in its own right: a Hilbert space where an experimenter can act and make observations, and in principle make predictions for those observations as well, within the constraints that would be imposed by the apparatus and how it acts in effective field theory.  Such sub-Hilbert spaces can be thought of as code subspaces: the set of observables of the experimenter is constrained to lie in the code subspace. At the technical level, the idea will be to first decompose the fields $\phi_i(x)= \vev{\phi_i(x)}_B + \sum_{\lambda} f_{i,\lambda}(x)a^\dagger_{i,\lambda}+ \sum_{\lambda} f^*_{i,\lambda}(x)b_{\lambda,i}$
into raising/lowering operators of approximate wavelength $\lambda$. We need to include the $b$ modes to allow for the possibility that the field $\phi_i$ is complex, otherwise we have $b\simeq a^\dagger$. For brevity, we will take the field to be real. We also need to require that the $a, a^\dagger$ approximately satisfy the Weyl commutation relations. To impose a cutoff, we state that the set of $\lambda$ is restricted. We also impose that $\ket B$ is annihilated by the lowering operators $a$. This second condition is what defines the state operationally to be effectively a coherent state.

We make use of the  modes of the fields $\phi_i(x)$ acting on $\ket B$ to generate new states
\begin{equation} 
\ket{(i_1, \lambda_1), \dots (i_k, \lambda_k) ;B} = \prod_{j}  a^\dagger_{i_j,\lambda_j} \ket B
\end{equation}
for some such collection of pairs $(i_j,\lambda_j)$. We can think of this state as the background state $\ket B$ with some finite number of cutoff respecting excitations turned on. We will call our cutoff $\Lambda$. Usually, we interpret $\Lambda$ as a UV cutoff in effective field theory around a ground state, so that energies (frequencies) $\omega$ of individual excitations are bounded above by $\omega\leq\Lambda$. Here, we are constrained so that our set of excitations collectively stay below $\Lambda$. The cutoff $\Lambda$ should not be  in general thought of as simply a fixed shortest wavelength, nor as just an upper bound on the energy. It can also be position dependent and dependent on the different modes. In the work \cite{Almheiri:2014lwa}, the cutoff is implicit in the sense that we do not form a black hole. In the work of Papadodimas and Raju, the cutoff is described by not having too many actions on the reference state \cite{Papadodimas:2013jku}. This is again an implicit cutoff.
 
 We will call the Hilbert space 
\begin{equation}
{\cal H}_{B, \Lambda}= Span(\ket{(i_1,\lambda_1),...(i_k,\lambda_k);B} \ | \{(i_1,\lambda_1),...(i_k,\lambda_k)\} \leq \Lambda  )
\end{equation}
the code subspace associated with the background $\ket B$ and the cutoff $\Lambda$. This will be sometimes abbreviated to ${\cal H}_{\rm{code}}$. 
This is in accordance with the definition of code subspace found in the work of Almheiry, Dong, Harlow \cite{Almheiri:2014lwa} on quantum error correction and it also matches the effective description 
of states generated from a reference black hole state in the work of Papadodimas and Raju \cite{Papadodimas:2013jku,Papadodimas:2015jra}. This also matches the definition of the nearby Hilbert space of states in our previous work \cite{Berenstein:2017abm}.
The advantage of using the language of code subspaces is that it makes three items automatic. First, it is a Hilbert space, so that we can do quantum mechanics inside ${\cal H}_{\rm{code}}$. Secondly, effective fields act simply on it. Finally, there is an explicit cutoff $\Lambda$, so this does not need to be repeated again and again: it is part of the definition of the code subspace itself.

If the state $\ket B$ is an  excited state (not a ground state), one can in principle find many states that have a similar energy to $\ket B$ but that are not generated in this way. One should think of the
code subspace ${\cal H}_{B, \Lambda}$ as the set of states that is accessible to an experimenter who can control the excitations of the fields $\phi_i$ below the cutoff. In this sense, this is the natural 
home for effective field theory. As an experimenter builds a better experiment, the cutoff might change and more states can become available. However, the effective field theory description might break down. This is not a failure of quantum mechanics, but of the simplified description of the Hilbert space of available states that the experimenter can access.

With this definition, the fields $\phi_i$ have been given to us, at the very least  in an implicit form,  as well as the mode expansion.  In general, we could expect that there are non-linear field redefinitions to worry about, as they might generate states that do not belong to the code subspace. We also have to worry that under time evolution the states might exit the code subspace. As long as we can stay comfortably inside $H_{\rm{code}}$ for some fixed amount of time we will be content. To do so we  will also include a temporal cutoff in the time during which experiments can be performed. The second problem  is not obviously an immediate issue if $\ket B$ is an energy eigenstate, but the problem will kick in as soon as we act on the state. 
We do not address these issues directly for general setups, rather, we will leave these issues implicit in the definition of $\Lambda$ itself, thinking of it as a set of all the necessary cutoff information.

One might think that this is overly pedantic. The purpose of this paper is to show that this structure is the only sense in which one can do effective field theory
in a particular subsector of a gravitational theory.  It will turn out that different code subspaces will 
generically be incompatible. That is, assume that a  state belongs to two such code subspaces  $\ket \psi \in  {\cal H}_{B,\Lambda}, {\cal H}_{B',\Lambda'}$. The topology of $\ket B$, $\ket {B'}$ and the number of (effective)  fields might differ substantially to the point where even though the state $\ket \psi$ is well defined, we cannot say what topology it has (the one of $\ket B $ or $\ket {B'}$) nor the number of fields. More importantly, one code subspace might recycle a field of another code subspace nonlinearly into many fields. What this will mean is that the physical answer to many  (interpretational) questions  can only be answered inside the different code subspaces, but not in the full Hilbert space ${\cal H}$.

Our goal in the rest of the paper will be to explain how to construct a particular collection of code subspaces explicitly, including the effective fields and the cutoff and to show precisely how they are incompatible.

\section{The action of traces on Young tableaux}\label{sec:YT}

We will now consider a particular set-up, where we can study effective field theory explicitly. The half-BPS states of ${\cal N}=4 $ SYM on the sphere are in one to one correspondence with the gauge invariant local operators that are build out of polynomials of a single scalar field $Z(x)$ (which we will take to be in the adjoint in the adjoint of $U(N)$). This space is converted into a Hilbert space via the operator state correspondence that is available in conformal field theories. We write the map as follows $\CO\to \ket \CO$.
For us to understand the Hilbert space of states, we need to determine the norms of states. The norms of states that correspond to local operators come the Zamolodchikov norm of the operator, obtained from the two point function as follows
\begin{equation}
\vev{\CO^\dagger(x) \CO(0)} = \frac{\braket {\CO}{\CO}}{|x|^{2\Delta_\CO}}
\end{equation}
What we need now is the complete list of operators in a basis that is suitable for computations. This problem was solved in \cite{Corley:2001zk}, where it was noted that a Schur polynomial basis was orthogonal
(this is based on the fact that characters of $Z$ in irreducible representations of $U(N)$ are orthogonal). However, there is another basis made of string states, which are traces , that is also useful and can be used to define the supergravity fields of $AdS_5\times S^5$. It is the traces that are used to define the extrapolate dictionary of $AdS/CFT$ \cite{Witten:1998qj}. Thus, it is necessary to study both basis to get to the complete physical description.

There are two natural ways to construct gauge invariant operators from a matrix $Z$. One of them is to take 
traces of powers of $Z$, $\Tr(Z^m)$ and to consider the set of linear combinations of multi-traces of $Z$. The other is to think of an $N\times N$ matrix $Z$ as an element of
$GL(N,\BC)$. Then we can take the character of $Z$ in some representation of the group $GL(N,\BC)$,  $R$, and denote the result as
$\chi_R(Z)$. The latter are classified by Young diagrams.

These two bases of gauge invariant operators generate the same linear space and can be related to each other algebraically. For example, the fundamental representation, with Young Tableau $\square$, can be related to $\Tr(Z)$ via its character as
\begin{equation}
\chi_{\square}(Z)= \Tr(Z).
\end{equation}

To write the relationship for other states, we need a few more definitions.
Let $[\sigma]$ be a conjugacy class of $S_n$. The conjugacy class of $[\sigma]$  is in one to one correspondence with group elements $\sigma$ of the same cycle decomposition, where there are $n_j([\sigma])$ cycles of length $j$, so that $n=\sum_j j n_j([\sigma])$.
See appendix \ref{sec:appa} for details on how the cycle decomposition is obtained from a group element.
To each such cycle, we associate the trace $\Tr(Z^j)$, so that to the element $[\sigma]$ we can associate the 
 monomial in the traces 
 \begin{equation}
 [\sigma]\to \prod_j (\Tr(Z^j))^{n_j([\sigma])}.
 \end{equation} 
If $R$ is represented by a Young diagram with $n$ boxes, which we indicate by $R_n$,  then
\begin{equation}
\chi_{R_n}(Z) =  \frac 1 {n!} \sum_{[\sigma] \in \rm{Conj}[S_n]} \chi_R([\sigma]) d_\sigma \prod_i (\Tr(Z^i))^{n_i([\sigma])}
\end{equation}
where $d_\sigma$ is the number of elements of the conjugacy class,   $\chi_R([\sigma])$ is the character of $\sigma$ in the representation of the group $S_n$ with the same Young diagram as 
$R_n$ (these are in one to one correspondence via Schur-Weyl duality). 
This explains how to write the basis $\chi_R(Z)$ in terms of traces. The map is invertible (the fact that the relationship between conjugacy classes and representations is invertible is true for any finite group, see  \cite{fulton2013representation}). The result of this inversion is 
\begin{equation}
(\Tr(Z^i))^{n_i([\sigma])}= \sum_{[R] \in \rm{Reps}[S_n]} \chi_R([\sigma^{-1}])\chi_{R}(Z)
\end{equation}

In what follows, we will define the variables $t_\ell= \Tr(Z^\ell)$, and we will label the representations of $R$ directly in terms of Young diagrams. The length of the cycle $\ell$ will be called the degree of $t_\ell$, and the number of boxes of a Young diagram $n$ will be the degree of the Young diagram. 
With this convention we have that $ [\sigma]\to \prod_i t_i^{n_i([\sigma])}$.
 The sum $\sum_\ell n_\ell \ell = n$, so we have that the degree of each of the monomials is equal to the degree of the Young diagram, and acting with an extra trace $\Tr(Z^\ell)$ will be multiplication by 
 $t_\ell$. Acting with $t_\ell $ on $\chi_R(Z)$ (by multiplication), will have degree $\deg(R)+\ell$ and can be expressed in terms of the basis of the $\chi_{\tilde R}$, with $\deg(\tilde R) = \deg(R)+\ell$.

 For example, we can take the state
\begin{equation}
(6,4,2,1) = \begin{ytableau}
\ & & & & &\\
& & &\\
&\\
\
\end{ytableau}
\end{equation}
where we label $R= (6,4,2,1) $ by the length of the rows of the Young diagram and we think of the Young diagram as the gauge invariant operator $\chi_{R}(Z)$ itself. Now, we want to act with one of the $t_\ell$, and see what linear combination of representation characters we get. The answer is actually simple.

We will do the particular example of $t_4 (6,4,2,1) $. Acting with $t_4$ will give us the following result
\begin{eqnarray} \ytableausetup{boxsize=1em}
t_4 \begin{ytableau}
\ & & & & &\\
& & &\\
&\\
\
\end{ytableau} &=& \begin{ytableau}
\ & & & & & & \bullet  & \bullet & \bullet & \bullet\\
& & &\\
&\\
\
\end{ytableau} - \begin{ytableau}
\ & & & & & &\bullet \\
& & & &\bullet  &\bullet &\bullet \\
&\\
\
\end{ytableau} \\ &&
-
\begin{ytableau}
\ & & & & &\\
& & & &\bullet\\
& &\bullet &\bullet &\bullet \\
\
\end{ytableau}
- \begin{ytableau}
\ & & & & &\\
& & &\\
& &\bullet &\bullet\\
\ &\bullet &\bullet
\end{ytableau} + 
\begin{ytableau}
\ & & & & &\\
& & &\\
&\\
\ &\bullet\\
&\bullet\\
\bullet &\bullet
\end{ytableau}\\
& & -
\begin{ytableau}
\ & & & & &\\
& & &\\
&\\
\bullet\\
\bullet\\
\bullet\\
\bullet\\
\end{ytableau}
\end{eqnarray}
where we have indicated with circles the extra boxes that are attached to the original Young diagram $R$.
The action on any given Young diagram is given by applying the following rules: 
\begin{enumerate}
\item The original Young diagram sits inside the added boxes. 
\item 
The set of new extra boxes are arranged in a pattern where they all touch each other and give rise to a proper diagram when combined with the original. 
\item The set of new boxes snake around the edge of the old diagram (this means that no square pattern set of $2\times 2$ boxes  can be found in the new boxes). Sets of boxes with this property are called skew-hooks.
\item 
The coefficients are all $\pm 1$. The sign is determined by how many rows the new boxes cover: $+1$ if the new boxes sit in an odd number of rows, and $(-1)$ if it is even. 
\item
The sum is over all possible ways of attaching a skew hook of the right length (in this case four, as we acted with $t_4$) to the original Young diagram.
\end{enumerate}
We can write these conditions as follows
\begin{equation}
t_\ell\;\chi_Y(Z) = \sum_{h\in\text{Skew hooks of length $\ell$}} (-1)^{H(h)-1} \chi_{Y+h}(Z)\label{eq:action}
\end{equation}
where the symbol $H(h)$ indicates the height of the hook (the number of rows it subtends). Also, the length of a skew hook, $|h|$, is the number of boxes it has.

As we have argued, the space of gauge invariant operators is endowed with a metric, the Zamolodchikov metric. This is a positive definite metric and is identified with the Hilbert space norm in the quantum theory on the cylinder. The norm  for each Young diagram state $\ket Y$ can be evaluated as follows. We first label the boxes of the diagram, adding one as we go to the right and subtracting one as we go down
\begin{equation}\ytableausetup{boxsize=2em}
(6,4,2,1) = \begin{ytableau} 
\ +0 &+1 &+2 &+3 &+4 &+5\\
-1 & 0 &+1 &+2\\
-2&-1 \\
-3 \
\end{ytableau}
\end{equation}

That is, to each box in position $(i,j)$ (the label $i$ refers to the column, and the label $j$ refers to the row of the box) we associate the number $i-j$. The norm is then computed as follows
\begin{equation}
\braket Y Y = \alpha ^{\# \text{boxes}}\prod_{(i,j) \in \text{boxes}} (N+i-j)
\end{equation}
where $\alpha$ is a normalization constant for the matrix field $Z$. Also, different Young tableaux are orthogonal. This was deduced in \cite{Corley:2001zk}.
 To simplify matters, we choose $\alpha= N^{-1}$. Then we have that 
\begin{equation}
\braket Y Y =\prod_{(i,j) \in \text{boxes}}\left(1+\frac{i-j}N\right)
\end{equation} 
so that the large $N\to \infty$ limit is simple and all the norms for each Young tableaux state are equal to one. 

Now, since the Young diagram states are orthogonal, we can consider a dual basis for the Young diagrams $\ket {\check Y'}$, so that we have the relation
\begin{equation}
\braket { \check Y ' }{ Y} = \delta_{Y,Y'}
\end{equation}
and it is easy to see that 
\begin{equation}
\ket {\check Y} = \ket{Y}/\braket YY 
\end{equation}
Using this dual basis, we can write the action \eqref{eq:action} as follows
\begin{equation}
\bra{\check Y+h} t_n \ket Y = (-1)^{H(h)} \delta_{|h|,n} 
\end{equation}
where again,  $|h|$ is the number of boxes in the skew hook $h$ and $\ket{Y+h}$ refers to a state $\ket{Y}$ with an added hook $h$. If we choose the basis to be orthonormal, we find that 
\begin{equation}
\bra{\check Y+h} t_n \ket Y =  \frac{\sqrt{\braket YY}}{{\sqrt{\braket{Y+h}{Y+h}}}}\bra{\hat Y+h} t_n \ket{\hat Y}
\end{equation}
where we are using $\ket{\hat Y}$ to represent an orthonormal state (as opposed to the state in the dual basis, which has a down check instead of a hat). So the action in the orthonormal basis is represented by 
\begin{equation}
\bra{\hat Y+h} t_n \ket{\hat Y} = (-1)^{H(h)} \delta_{|h|,n} \frac{\sqrt{\braket{Y+h}{Y+h}}}{\sqrt{\braket YY}}=(-1)^{H(h)} \delta_{|h|,n} \prod_{(i,j)\in {\text{boxes of h}}} \sqrt{\left(1+\frac{i-j}N\right)}
\end{equation}
And the adjoint action is
\begin{equation}
\bra{\hat Y-h} t_n^\dagger\ket{\hat Y} = (-1)^{H(h)} \delta_{|h|,n} \frac{\sqrt{\braket{Y}{Y}}}{\sqrt{\braket{ Y-h}{Y-h}}}=(-1)^{H(h)} \delta_{|h|,n} \prod_{(i,j)\in {\text{boxes of h}}} \sqrt{\left(1+\frac{i-j}N\right)}\label{eq:normalized}
\end{equation}

With this formula, we now have  the main computational tool we need for the rest of the paper. It should be noted that if we take the limit where $i,j$ are finite and $N\to \infty$, the set of Young tableaux states all have trivial norm and the coefficients for the action of the traces are all $\pm 1$. In this case, the half BPS states are described exactly by a $c=1$ left-moving chiral boson in $1+1$ dimensions. 

\section{Defining the code subspaces for concentric configurations.}\label{sec:geomcode}

Let us consider a particular case of an LLM geometry that is  time independent and that is therefore an eigenstate of the Hamiltonian of the ${\cal N}=4 $  SYM theory on the $S^3\times \BR$ boundary. 
The LLM geometries are  described by  droplet configurations on the plane, and as they are time evolved they rotate uniformly about an origin in the LLM plane. The ground state is described by a disk, and the rotation center is located exactly at the center of the disk. For another configuration to be similarly time independent, the droplet configuration must be invariant under rotations around such an origin. This results in 
a droplet configuration that is described by a set of concentric rings. Because the configurations are time independent, they have an extra isometry symmetry in the supergravity description. This extra symmetry is exactly the rotation of the configuration around the origin. This is a diffeomorphism that does not vanish at infinity and is realized as a proper symmetry of the configuration. This is the symmetry associated with either time translation or to being an eigenstate of the  $R$-charge. 

\begin{figure}[ht]
\begin{center}
\includegraphics[width=7cm]{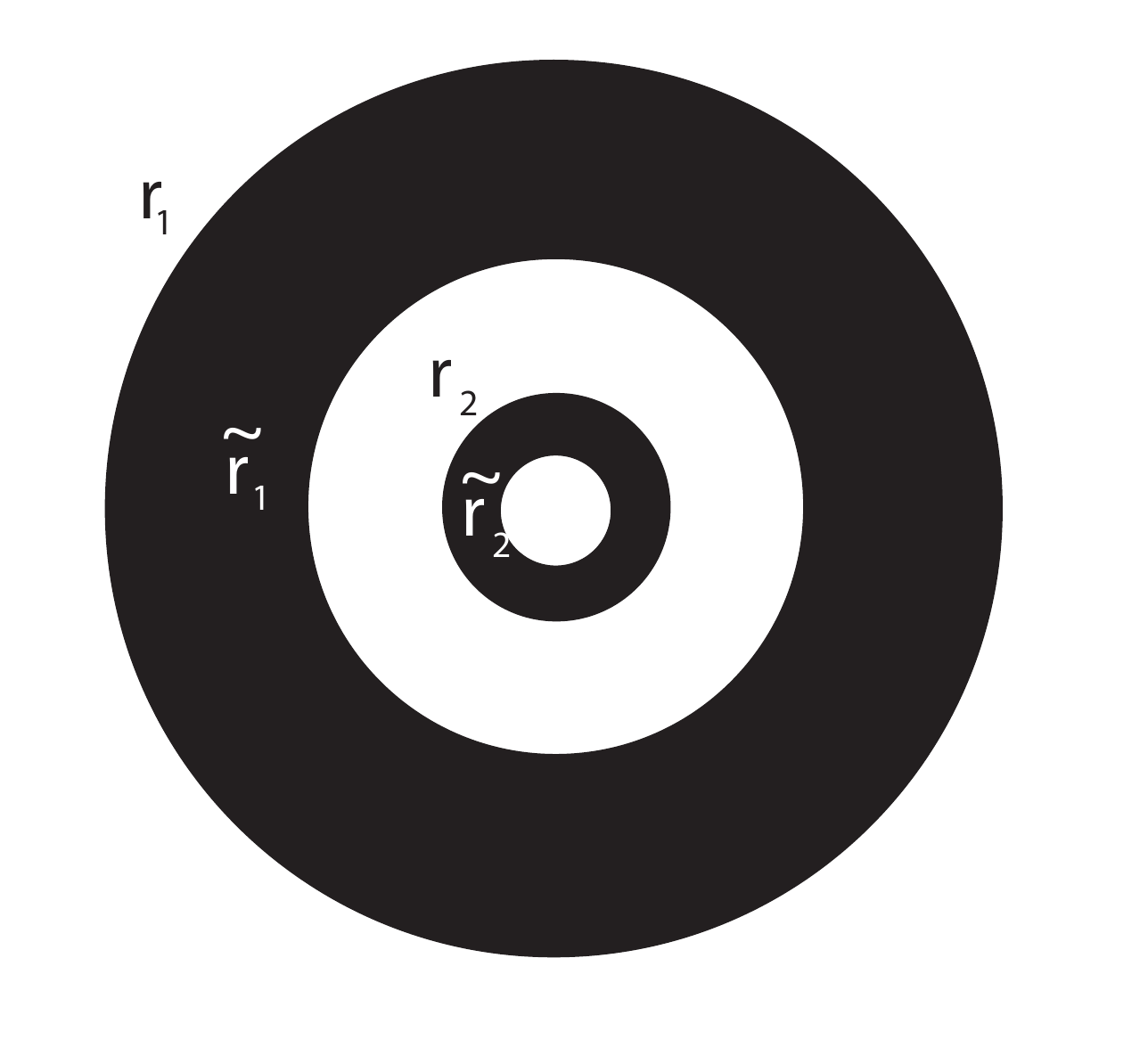}\caption{A circularly invariant LLM geometry. The radii are labeled by $r,\tilde r$, depending on if they go from black region to 
white region, or viceversa, starting from the outside and going inwards.} \label{fig:concentric}
\end{center}
\end{figure}

A particular example of a geometry is visualized in figure \ref{fig:concentric}. The radii of the boundaries are labeled as follows $r_1>\tilde r_1>r_2>\tilde r_2 >\dots $, from the outermost boundary inwards. We will call these edges and anti-edges, depending on if they are labeled with an $r_i$ or a $\tilde r_i$, that is, if they go from black to white (edges) or if they go from white to black (anti-edges) when tracing a straight line from the origin. This is in accordance with the convention established in \cite{Berenstein:2017abm}.
The important geometric parameters are the radii $r_1, \tilde r_1, \dots$ which are necessary to uniquely identify the geometry. We will assume that we are working at fixed large $N$, and that the disk representing the ground state has been normalized to have radius equal to one. We will also assume that the $r_i, \tilde r_i$ are of order one. The area of the black region is given by
\begin{equation}
A= \pi \sum_i( r_i^2 -\tilde r_i^2) = \pi
\end{equation}
which is the same as the area of the unit disk. This gives us one relation between the radii.

 The energy of the state (geometry) is given by 
\begin{equation}
E \propto N^2\left( \sum_i r_i^4 -\tilde r_i^4\right )-N^2
\end{equation}

\begin{figure}[ht]
\begin{center}
\includegraphics[width=7cm]{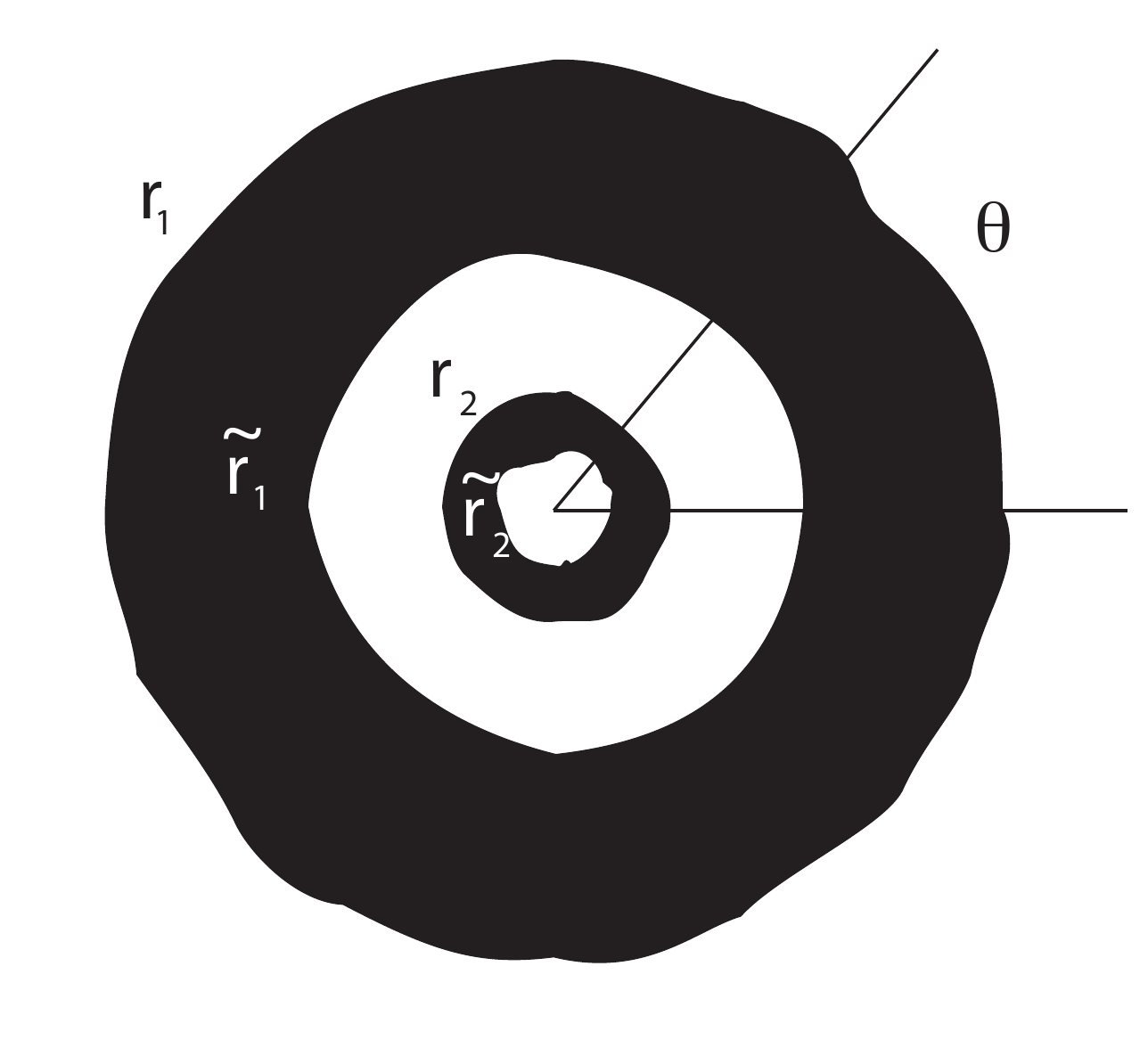}\caption{A  slightly deformed circularly invariant LLM geometry. The radii are now given by $r_i(\theta),\tilde r_i(\theta)$} \label{fig:concentric2}
\end{center}
\end{figure}

As long as the radii are moderately spaced, the solution is weakly curved and effective field theory is valid. Small deformations of the geometry that preserve the supersymmetry can be  characterized by 
having the $r_i, \tilde r_i$ vary with the angle around the origin as follows $r_i(\theta)= r_i +\delta r_i(\theta)$, $\tilde r_i(\theta) =\tilde  r_i +\delta \tilde r_i(\theta)$ with $\delta r, \delta \tilde r<<1$
and more precisely, we require that the configuration is fairly smooth so that the wiggles are not too pronounced. This is what we mean by long wavelength fluctuations. This is depicted in figure \ref{fig:concentric2}.
Larger deformations will have either short wavelength (rougher edges), or larger amplitude resulting in more pronounced creases.

From the point of view of supergravity, it is obvious that to each of the  radii $r_i, \tilde r_i$ we can associate a function of one variable $\theta$ that preserves the supersymmetry of the configuration and that therefore the effective field theory -- restricted to the half-BPS states-- is now described by 
many functions of $\theta$. Since each $r_i, \tilde r_i$ can be in principle deformed independently of the others, we have to associate an effective field with each such (anti-) edge. Each such field should result 
in a left chiral field $\phi_i(\theta), \tilde \phi_i(\theta)$, just like the single edge of the ground state results in such a field. 

In the limit where the $\delta r_i, \tilde \delta r_i <<1$, we can expand the area of the regions to linear order in these deformations. We get that the areas of the regions are given by one of the two expressions (the left for black regions and the right for white)
\begin{equation}
A_i= \frac 12 \int d\theta(r_i^2(\theta)-\tilde r_i^2(\theta)), \tilde A_i =\frac 12 \int d \theta( \tilde r_i(\theta)^2-r_{i+1}(\theta)^2)
\end{equation}
and to leading order in fluctuations we get that the variations in the area are given by  
\begin{equation}
\delta A_i=  \int d\theta(r_i \delta r_i (\theta)-\tilde r_i \delta\tilde r_i (\theta)), \delta \tilde A_i = \int d \theta ( \tilde r_i \delta \tilde r_i(\theta)-r_{i+1}\delta r_{i+1}(\theta))
\end{equation}
 Because the areas are quantized in natural units due to Dirac quantization condition, we need to require that the $\delta A_i = \delta \tilde  A_i =0$, and this implies that 
 \begin{equation}
 \int d\theta \delta r_i(\theta)= \int d \theta \delta \tilde r_i(\theta)=0
 \end{equation}
We infer that the fluctuations have a Fourier expansion with the zero mode  missing for each fluctuation. 

For quantization, we also need the Poisson bracket between the $\delta r_i(\theta)$. This was calculated in \cite{Maoz:2005nk}. One can derive it from the Hamiltonian and locality in $\theta$. Basically, we need that
\begin{equation}
\{ H, \delta r_i(\theta)\} = \dot r_i(\theta) = -\partial_\theta r_i(\theta) 
\end{equation}
and similarly for $\delta \tilde r_i$.
Since the energy function is the Hamiltonian, the Poisson bracket is as follows
\begin{equation}
\{\delta r_i(\theta), \delta r_j(\phi)\} = \frac {\delta_{ij}}{N^2 r_j(\theta)^2} \partial_\phi \delta(\theta-\phi)
\end{equation}
whereas for the other variables we find a sign change
\begin{equation}
\{\delta \tilde r_i(\theta), \delta \tilde r_j(\phi)\} = - \frac {\delta_{ij}}{N^2 \tilde r_j(\theta)^2} \partial_\phi \delta(\theta-\phi)
\end{equation}
that follows because the Hamiltonian for the $\delta \tilde r$ is actually negative definite.  The cross term vanishes.

We interpret the prefactor in front of the derivative of the delta function as an effective notion of the Planck constant $\hbar$ for the corresponding background field determined by the geometric data.  
As is usual in the AdS/CFT correspondence, $\hbar$ scales as $1/N^2$, and this is the  normalized Newton constant \footnote{ Notice that if we scale $r, \delta r$ by the same scale factor to remove the $1/N^2$ pieces, we find that the normalized value of the radius $R$ then scales as $R \simeq \sqrt N$, and the area of the disk is proportional to $N$.}.

With these conventions, the canonically normalized fluctuating fields $\phi^C$, and $\tilde \phi^C$ depend on the values of the geometric parameters $r_i, \tilde r_i$, as follows
\begin{equation}
\phi^C_i (\theta) =  N r_i   \delta r_i(\theta) = \frac N 2 \delta (r_i^2(\theta))
\end{equation}
and similarly for $\tilde \phi^C$. The Fourier modes of the $\phi^C, \tilde \phi^C$ will have canonical commutation relations, such that 
\begin{equation}
\phi^C_i (\theta)= \sum_n \phi^C_{i,n} \exp(i n \theta)
\end{equation}
and 
\begin{equation}
\{ \phi^C_{i,n} , \phi^C_{j,m}\} = n \delta_{ij}\delta_{n,m} 
\end{equation}
whereas for the $\tilde \phi$ we get a sign change
\begin{equation}
\{\tilde  \phi^C_{i,n} ,\tilde  \phi^C_{j,m}\} =- n \delta_{ij}\delta_{n,m} 
\end{equation}
One can then show that the $\phi^C_{n},\tilde \phi^C_n$ are either raising or lowering operators with energy $n, -n$ respectively (the $\tilde C$ fluctuations reduce the energy).

Now that we have our canonical mode fields, we can define the code subspaces as  in section \ref{sec:eff}. We just take the concentric ring background configuration and act with the raising operators $\phi^C_{-n}, \tilde \phi^C_m$ a finite number of times, with a cutoff on $n,m$ and the number of raising operators acting on the reference state, which is specified by the radii, and the quantization condition that all the lowering modes of the effective edge fields are in their ground state. The precise details of the cutoff are yet to be specified, and the result should be understood to be a leading order approximation in a large $N$ expansion, so there might be $1/N$ corrections that need to be studied more carefully.

Notice that the definition of the code subspace is fairly straightforward, but the determination of the effective modes took some work. It also assumes a particular action of the modes of the fields on the reference state and by relying heavily on the classical analysis in gravity, we do not have a direct access to how the cutoff should be correctly implemented.

A second issue that needs attention is to make sure that the code subspace that we have defined this way is compatible with the holographic boundary operator actions. That is, we want to show that acting with the operators that preserve the supersymmetry and that are realized in the boundary does not take us out of the code subspace.
To do this, we need to notice that the boundary operators measure the multipole moments of the droplet distribution. These can be written as follows
\begin{equation}
\phi_n \simeq \int \rho(r,\theta) \left[r \exp(-i \theta)\right]^n r d r d \theta
\end{equation}
where $\rho$ is the region of the plane that is filled. After some manipulations where we do the radial integral first \cite{Berenstein:2016mxt}, these modes are written as follows
\begin{equation}
\phi_n = \frac{1}{ n+2} \int  \sum_i\left[r_i(\theta) \exp(-i \theta)\right]^n r_i^2(\theta)-\left[\tilde r_i(\theta) \exp(-i \theta)\right]^n \tilde r_i^2(\theta) d \theta
\end{equation}
To linearized order we have that 
$r_i(\theta) = r_i +\delta r_i(\theta)$, so we find that
\begin{equation}
\phi_n =  \int  \sum_i \left[r_i^{n+1} \delta r_i (\theta) \exp(-i n \theta)-\tilde r_i^{n+1}\delta r_i (\theta) \exp(-i n\theta) \right] d \theta
\end{equation}
which in terms of the canonical fields becomes
\begin{equation}
\phi_n= N^{-1} \sum_i \left[r_i^{n} \phi^C_{i,n} -\tilde r_i^{n}\tilde \phi^C_{i,n}  \right]
\end{equation}
The factor of $1/N$ in the prefactor is to be thought of as $\sqrt \hbar$, which is the standard size for quantum fluctuations.  In this sense, we should remove it and the normalized operator for boundary insertions should be given by
\begin{equation}
\hat \phi_n= \sum_i \left[r_i^{n} \phi^C_{i,n} -\tilde r_i^{n}\tilde \phi^C_{i,n}  \right] \label{eq:PartBog}
\end{equation}
With these conventions we have that for the ground state where $r_1=1$ and all other $r_i, \tilde r_i$ vanish, we have that 
\begin{equation}
\hat \phi_n= \phi^C_{1,n}
\end{equation}
Since the field $\hat \phi$ is in general clearly a linear combination of operators in the code subspace, it  belongs to the code subspace. To go beyond linearized order, we need to normal order the expressions. The non-linear terms that are generated will be polynomials in $\phi^C_{i,n}$ and $\tilde \phi^C_{i,n}$ suppressed by additional powers $1/N$. 

What we want to do now is reproduce these same results without relying on the semi-classical description, but directly in terms of the matrix variables. What is important for us is that the concentric circle configurations have a simple description in terms of Young tableaux \cite{Lin:2004nb}. Therefore it is possible to analyze the physics of the cutoff in terms of the trace variables as in section \ref{sec:YT}.
This will permit us to describe the cutoffs better and to verify directly the expression \eqref{eq:PartBog}. This can be understood as a test of the LLM geometry map. We will tackle this problem in the next section.

\section{Code subspaces in the Young tableaux formalism}\label{sec:codeyoung}

As is by now well understood, concentric ring classical configurations in the LLM plane correspond to Young tableaux with only a few corners (see \cite{Mosaffa:2006qk} for more details). A typical such Young tableaux looks as portrayed in the diagram \eqref{eq:LMlabels}
\begin{equation}
\begin{ytableau}
\ & & & & & \none[ \dots] &M_1 \\
& & & & &\none  &\none[\vdots] \\
& & & & & \none[\dots] &L_1\\
\ & & \none[\dots] &  M_2 \\
& &\none &\none[ \vdots] \\
&&\none[\dots]& L_2\\
\none[\vdots]
\end{ytableau}\label{eq:LMlabels}
\end{equation}
where the length of the long rows are of size $M_1, M_2, \dots$, and the depth of the columns is $L_1, L_2, \dots$.
With these conventions, the empty corners to the right of the $M_i$ (concave corners) have coordinates given by
\begin{equation}
(i,j) \in \{ (M_1+1,1), (M_2+1, L_1+1), \dots (M_k+1, L_{k-1}+1) \}
\end{equation}
Similarly, the convex corners of the edge of the tableaux are given by the coordinates
\begin{equation}
(i,j) \in \{(M_1, L_1) \dots (M_k, L_k)\}
\end{equation}
We will call this state the reference state $\ket \Omega$. It is around this state that we want to build an effective field theory of the LLM states that mirrors the gravity construction.
To such a tableaux with widely spread out corners we can associate a Hilbert space of small fluctuations. These are additional small tableaux that can be attached to each corner. In the young diagram depicted in
\eqref{eq:add_subcorner}, we see an example of adding a small tableaux to the concave corner depicted with the symbol $+$ and also a tableaux that is substracted from the convex corner 
and depicted with the symbol $-$. This idea was originally sketched in \cite{Koch:2008ah}, but was not fully realized at the time. It was implemented in the strict $N\to \infty$ limit in \cite{Berenstein:2017abm} and better estimates for various quantities  were obtained in \cite{Lin:2017dnz}.

 Here we have a full implementation of the details at finite $N$.
\begin{equation} \ytableausetup{boxsize=1em}
\begin{ytableau}
\none[\ddots]&\none&\none&\none&\none&\none&\none&\none&\none& &+&+\\
\none&\none&\none&\none&\none&\none&\none&\none&\none&&+\\
\none&\none&\none[\ddots]&\none&\none&\none&\none&\none&\none&\\
\none&\none&\none&\none&\none&\none&\none&\none&\none&\\
\none&\none&\none&\none&\none[\ddots]&\none&\none&\none&\none&\\
\none&\none&\none&\none&\none&\none]&\none&\none&\none&\\
\none&\none&\none&\none&\none&\none& \none &&&\\
\none&\none&\none&\none&\none&\none&\none& &&- \\
&&&&&&&&-&-
\end{ytableau}\label{eq:add_subcorner}
\end{equation}
It is easy to see that we can define a small Hilbert space for each corner. This Hilbert space is the set of small Young tableaux (whose sides are much smaller than the sides of the big tableaux with few corners). There will generically be two types of corners: the ones that have $+$ boxes, and the ones that have $-$ boxes. These correspond to the two types of corners of the reference tableaux. For convenience, we will label them with the $(i,j)$ values of the first corner that we can add or substract, and each of these corner Hilbert spaces will be called ${\cal H}_{(i.j)}$.
By construction, we find the small Hilbert space of states relative to the reference state $\ket \Omega$, which we will call the code subspace, can be  decomposed  as follows
\begin{equation}
{\cal H}_{\rm{code} \ket \Omega}= \prod_k {\cal H}_{(M_k+1, L_{k-1}+1)}\otimes  \prod_k {\cal H}_{(M_k, L_{k})}
\end{equation}
As of yet, we have not specified the size of the factors of code subspace. We will proceed to do this later. What we need to do right now is to understand in a little more detail the 
${\cal H}_{(i.j)}$ factors. The idea is that each of these is characterized by a Young diagram. 
There are two cases to consider: the $+$ subspaces and the $-$ subspaces. 

Let us begin with the $+$ subspaces. These are labeled by ${\cal H}_{(M_k+1, L_{k-1}+1)}$. What we are interested in to begin with are the factors associated with adding and substracting boxes, as in equation \eqref{eq:normalized}. We will use new sets of relative labels to the reference corner  $(\Delta_i,\Delta_j)=(i-M_k,j-L_{k-1})$. In this way the square root factors from before read
\begin{equation}
\left(1+\frac{i-j}N\right)^{1/2} \to\left(1+\frac{M_k-L_{k-1}}N+\frac{\Delta_i-\Delta_j}N\right)^{1/2} \simeq \left(1+\frac{M_k-L_{k-1}}N\right)^{1/2}\label{eq:simpli}
\end{equation}
in the limit where $N$ is large and the $\Delta_i, \Delta_j$ are of order one.

When we add a skew hook with $s$ boxes to a Young diagram that belongs  ${\cal H}_{(M_k+1, L_{k-1}+1)}$, we would associate the factor
\begin{equation}
\left(1+\frac{M_k-L_{k-1}}N\right)^{s/2}
\end{equation}
and we need to identify this with an action as we would have in equation \eqref{eq:PartBog}. The correct identification to have a match is that 
\begin{equation}
r_k= \left(1+\frac{M_k-L_{k-1}}N\right)^{1/2}
\end{equation}
With this, we find that 
\begin{equation}
r_k^2 =1+\frac{M_k-L_{k-1}}N
\end{equation}
so that the $M_k, L_{k-1}$ are clearly geometric. To have $r_k$ of order one, we need $M_k, L_{k-1}$ to be of order $N$. For convenience, we add $L_0=0$ so that the uppermost corner  can be treated uniformly with the others.

The idea now is that to each such corner we will assign a set of variables $ t_{k,\ell} := t_{M_k, L_{k-1},\ell}$ labeled by an integer $\ell$, such that they act as traces in the small Young tableaux alone. That is, we write is as follows
\begin{equation}
\bra{\hat Y_k+h} t_{k,n} \ket{\hat Y_k} = (-1)^{H(h)} \delta_{|h|,n}\label{eq:subYT}
\end{equation}
where $Y_k$ is the small Young tableaux in the corner.

Now we need to do something similar with the convex corners. The beginning setup is the same, starting at the $(M_k,L_k)$ corner  but now we are substracting boxes. The relative coordinates will now be given by
\begin{equation}
(\Delta_i, \Delta_j)= (L_k-j,M_k-i)
\end{equation} 
so that they are both positive.
Notice that we have switched the $i,j$ labels in the definition of the left. With this convention we get that the corresponding square root factor is still of the form
\begin{equation}
\left(1+\frac{i-j}N\right)^{1/2}= \left(1+\frac{M_k-L_k}N + \frac{\Delta_i - \Delta _j}N\right)^{1/2}{\substack{\longrightarrow\\N\to \infty}} \left(1+\frac{M_k-L_k}N \right)^{1/2}\label{eq:simpli2}
\end{equation}
where both $\Delta_i$ and $\Delta_j$ appear with the same sign as before. Notice that $\Delta_i$ increases as we go up the diagram, and $\Delta_j$ increases as we go to the left.
This way we find that 
\begin{equation}
\tilde r_k^2 = 1+\frac{M_k-L_{k}}N
\end{equation}
What this means is that the vertical direction in the $-$ tableaux should be thought of in a similar way to the horizontal direction in a $+$ tableaux, and similarly the horizontal direction in the $-$ tableaux should be thought of as the vertical direction in a $+$ tableaux. That is, the conventions for the tableaux are reflected.
We now want to introduce $\tilde t_{k, \ell}$ variables that act only in convex corners.
To get an equation that works as \eqref{eq:subYT}, we need to modify it to look as follows
\begin{equation}
\bra{\hat Y_{\tilde k}+h} \tilde t_{k,n} \ket{\hat Y_{\tilde k}} = (-1)^{W(h)} \delta_{|h|,n}\label{eq:subYTtrans}
\end{equation}
where instead of measuring the height of the skew hook, we measure the width of the skew hook in the $-$ boxes. 
 For hooks with an odd number of boxes, the vertical and horizontal parity coincide. 
Whereas for skew hooks with an even number of boxes, they are opposite. This means that relative to the usual conventions, we have set up $\tilde t_\ell$ to act as $-t_\ell$ for $\ell$ even. It is more convenient to have $\tilde t_{\ell}$ to have a uniform negative sign in all actions. This is done by changing signs in the definition of $Y_{\tilde k}\to (-1)^{\# \rm{boxes}} Y_{\tilde k}$. That way both even and odd $\tilde t_{k,\ell}$ act with a minus sign relative to the usual convention. 

We can now ask how $t_\ell$ acts on a state in the code subspace. 
It is straightforward to show that in general we can write
\begin{equation}
t_\ell \simeq \sum_k \left(r_k^\ell \, t_{k,\ell} - \tilde r^\ell_k \, \tilde t^\dagger_{k, \ell}\right)\label{eq:partbogcs}
\end{equation}
which is an equation that seems identical to equation \eqref{eq:PartBog}. The minus sign for the $\tilde t$ variables is the minus sign that we just introduced. This is part of the definition of how the $\tilde Y$  diagrams should be understood.

Because the different $t_k, \tilde t _k$ act on different subfactors of ${\cal H}_{\rm{code}}$, they automatically commute. Moreover, they also commute with each other's adjoints. The only non trivial commutation relations are between $t_k$ and their own adjoint, or between $\tilde t _k$ and their adjoints.
To get a good match we need to show that the $t_k, \tilde t_k$ variables should have canonical commutation relations. This was proven in our previous work \cite{Berenstein:2017abm} for tableaux without restrictions. Since the tableaux are restricted in size, this cannot be true for general states. After all, the representation of a harmonic oscillator algebra is always infinite.
This is the first formal cutoff we encounter. The commutation relations that we need 
\begin{equation}
[t^\dagger_{k, \ell}, t_{k, m}] = \ell \, \delta_{\ell, m}
\end{equation}
should be valid inside the factor of the code subspace, but {\em only when sandwiched between states in the code subspace}. If the $t_{k,\ell}$ take us out of the code subspace, then
we need to define their action. The bounds are implicit in that the small tableaux have small sizes, and the definition of their limits is still to be determined more carefully.
Here we see that the language of the code subspace is helping us to understand that the commutation relations we need are valid in a restricted subspace of the Hilbert space, and they can be arbitrary outside. The language of these relations automatically assumes that we are inside the code subspace. This is also the way the code subspaces do their  work in \cite{Almheiri:2014lwa}.

Now, by construction we have that multiplication by the $t_{k,\ell}, \tilde t_{k,\ell}$ act as raising operators in the small factors. Since $t_{\ell}$ adds boxes and $\tilde t_{k,\ell}$ substracts them, $\tilde t_{\ell}$
is more similar to $t_{\ell}^\dagger$. That is why we need to write the equation \eqref{eq:partbogcs} with daggered operators for the $\tilde t$ variables.

For convenience, since multiplying by $t_{k,\ell}$ is like a raising operator and their adjoint is like a lowering operator, we will rewrite the equation in a more standard Fock space language. We do this by stating that
\begin{eqnarray}
t_{k,\ell} &\to& b_{k,\ell}^\dagger\\
\tilde t_{k, \ell} &\to& c_{k, \ell}^\dagger\\
t_{k, \ell}^\dagger&\to& b_{k,\ell}\\
\tilde t_{k, \ell}^\dagger&\to& c_{k,\ell}
\end{eqnarray} 
These identifications are valid inside the code subspace. The $b,c$ oscillators have canonical commutation relations.
The action of $t_{k,\ell}$ becomes
\begin{equation}
t_\ell = \sum_k \left( r_k^\ell b^\dagger_{k,\ell} - \tilde r^\ell_k c_{k, \ell}+O(1/N) \right) \label{eq:partbogosc}
\end{equation}
and we ignore the $1/N$ corrections when we match to supergravity. Now it is clear that \eqref{eq:partbogosc} is identical in form to \eqref{eq:PartBog}. Where the field modes have canonical commutation relations, just like the supergravity modes do.
This implements the requirements of equation \eqref{eq:PartBog} exactly. That is, the code subspace in the Young tableaux basis can be put into correspondence exactly with the code subspace in supergravity. 

Moreover, we have seen that there is an implicit cutoff on the size of the small tableaux. This is not immediately apparent in the supergravity construction where one is formally taking the limit  $N\to \infty$ first. To argue for the cutoffs, one needs to follow \cite{McGreevy:2000cw} and argue that a type of stringy exclusion principle (similar to \cite{Maldacena:1998bw}) is responsible for a cutoff on the number of modes and their amplitudes. 
To proceed further, we need to understand the implicit cutoffs explicitly and explore the physics that is {\em beyond classical  supergravity}.

\section{Cutoff physics in the Young tableaux formalism} \label{sec:cutoff}

The first step in the process of understanding the cutoff is to describe when the linearization implied by equation \eqref{eq:partbogcs} is correct.  In essence, we want to understand how  when multiplying by $t_\ell$, the subleading terms in $N$ in the expression \eqref{eq:simpli} or \eqref{eq:simpli2} accumulate when we vary $\ell$ and take $\ell$ large. This limit will give us a UV cutoff on the effective modes beyond which non-linearities matter. 

The idea is that the products on each skew hook will be of the form
\begin{equation}
|\bra{\hat Y_k+h} t_\ell \ket{\hat Y_k}|=\prod_{k=\alpha+1}^{\alpha+\ell}\left(r^2_m + k/ N \right)^{1/2}= r_m^{\ell} \exp( \sum_k \log (1+ k/(N r^2_m)))
\end{equation}
and we will look at cases where $r_m$ is of order $1$ and $k<<N$. The term in the exponential  can be further approximated by
 \begin{equation}
\sum_k  k/(N r^2_m))= O( \ell^2/( N r_m^2))
 \end{equation}
We want these corrections to be small for each skew hook, which means that we want in general $ \ell^2/( N r_m^2)<< 1$. That means that we should have the label $\ell$ scaling at most as $\ell< \epsilon N^{1/2}$ where $\epsilon$ is a small number (this is the same scaling that is observed in studies of the BMN string \cite{Kristjansen:2002bb}, that ends up being a special case of the LLM geometries: the vacuum geometry). For us it is a choice that tells us how big of an error we should allow. A similar (slightly weaker) limit is obtained from three point functions \cite{Garner:2014kna} (see also \cite{Dhar:2005su} and references therein for earlier work on the exact three point functions).

That is, the code subspaces associated with the corners have a bound on the size of skew hooks. We also want the bound to apply to excited states, so all the $\Delta_i$ and $\Delta_j$ should also fit in this bound. In essence, the allowed Young tableaux on each corner is essentially a tableaux that fits in a square of order $\sqrt N\times \sqrt N$ around each corner.
If the $L_i, M_i$ are well separated from each other, the different tableaux on each corner cannot interfere with each other, because the horizontal or vertical difference between the corners 
is of order $N$. Since such a small tableaux has energy that is equal to the number of boxes in the tableaux, this means that the excitation energy above (below) the reference state is bounded and of order at most $N$. This is subleading in the supergravity 
description, because the energy of the supergravity solutions is of order $N^2$, but it is also a typical energy of a single giant graviton who scale is of order the AdS radius. This limit where the linear structure starts breaking down is due to $1/N$ corrections given by interactions of the local string excitations (as perceived by the extrapolate dictionary). 

Remember that the Planck scale quanta are associated to  energies of order $N^{1/4}$. An energy of order $N^{1/2}$ is roughly the energy of a Planck sized object that has been boosted by an ultra-relativistic factor of $\gamma\simeq N^{1/4}$. This means that the physics of these modes does not break down at the Planck scale, but at much higher energies and the notion of (local) Lorentz invariance for single particle states should be well respected at energies of order the Planck scale itself. Since the total size of the circle associated with the edge of the droplet is of order $N^{1/4}l_P$,  the Lorentz contraction obtained from a  boost of $N^{1/4}$ gives an effective circle of size $\ell_P$ (similar boost arguments have been used to describe matrix black holes \cite{Banks:1997hz}). In essence, the physics is breaking down when for a boosted object at the Planck scale, the Lorentz contracted circle on which it is moving is of order the Planck length.

Our description of the cutoff is that the allowed Young tableaux need to fit inside a square of size $w_i\times w_i$ where each $w_i$ scales as $N^{1/2}$. We can restrict the action of the $b,c $ modes so that if a skew hook falls outside these squares we get zero.  This would modify the canonical commutation relations between these modes only for tableaux that are nearly filling the allowed squares. Also, the restriction in depth is similar to the restriction that representations for $SU(M)$ vanish if their associated Young tableaux have a column of length larger than or equal to $M$. This restriction makes traces of length larger than or equal to $M$ dependent non-linearly on the smaller traces (these are the Mandelstam relations and they are closely related to the Cayley Hamilton relation, see for example \cite{Berenstein:1993gb}). The additional $b,c$ modes become (non-linearly) redundant when we hit this bound. The maximal bound on $\ell$ for each of the $b,c$ modes is fixed by the size of the square regions, and all small Young tableaux states can be generated from the action of $b,c,$ modes with these cutoffs (as long as we act by zero when we get out of the confining boxes).

\section{Uncertainty and entropy}\label{sec:uncent}

In our previous work we argued that in the strict $N\to \infty$ case one could calculate the  topology of LLM geometries by measuring the uncertainty and the entropy of the mode expansion for the $t_\ell$ actions on the corresponding Young tableaux state. Our purpose now is to understand how the answer changes when we take $N$ finite, or more precisely, when we take $N$ to be very large and the $L,M$ scale with $N$. In this way, we can take equation \eqref{eq:PartBog} or equivalently \eqref{eq:partbogosc} and compute the uncertainties for the actions by traces.
The result is very simple, by using Wick's theorem in the $b,c$ oscillators (this is the original technique  we used in \cite{Berenstein:2016pcx}, with the understanding that the tails of distributions contribute a very small amount). This is combinatorially equivalent to computing directly with the Young tableaux \cite{Berenstein:2017abm, Lin:2017dnz}, and  we get the following answers
\begin{eqnarray}
\vev{t^\dagger_s t_m}_{\Omega}&=& s \delta _{sm} \sum_i r_i^{2m} \equiv s \delta_{sm} S_m\\
\vev{t_m t_s^\dagger}_{\Omega}&=&s \delta _{sm} \sum_i \tilde r_i^{2m}\equiv s \delta_{sm} \tilde S_m
\end{eqnarray}
for the reference state $\ket \Omega$. 

For the previous case studied by us, the answers are given by specializing to $r_i=\tilde r_i=1$ for all $i$, in which case we would immediately get the number of edges and anti edges
by computing these expectation values. 

What we see in this case is that now the answer on the right hand side is geometric. We get  an algebraic sum of the powers of the $r_i$, or the powers of the $\tilde r_i$.
These are symmetric functions of the $r_i$ or the $\tilde r_i$ respectively.
If there is a finite number of these given by $N_{\rm{edges}},N_{\rm{anti-edges}}$,  we will find that there are algebraic relations between them. The first order for a non-trivial relation will be exactly when we have enough variables on the left hand side above to be able to compute the $r_i$ by solving for the roots of  a polynomial. To obtain the coefficients $A_i$ of the polynomial equation from the $S_m$, one uses Newton's equations given as follows
\begin{eqnarray}
A_1+S_1&=&0\\
2A_2+S_1 A_1+S_2&=&0\\
&\vdots &
\\
n A_n +S_1 A_{N-1}+\dots + S_{N-1} A_1+S_n&=&0
\end{eqnarray}
Once we pass the point where we have saturated the number of the different $r_i$, the corresponding $A_n$ will vanish: the rest of the putative $r_i,\tilde r_i$ would vanish \footnote{The $S_i$ are also Schur polynomials related to the totally antisymmetric representation for a matrix with eigenvalues $r_i$ or $\tilde r_i$.}.

In this sense, it is possible to get the topology for a concentric LLM geometry. One can follow a similar argument for coherent states of the $b,c$ oscillators, only as long as the notion of coherent states fits nicely within the cutoffs of the code subspace that were discussed in the previous section. In this sense, the deformations away from circularity of the droplets are of subleading order  in $N$. The easiest way to see this is that a classical shape deformation should typically cost an energy of order $N^2$, but the cutoff windows we have discussed  only allows for changes in energy of order $N$. 

Going a little bit further in comparison to our previous work, we notice that as long as we ignore the cutoffs, we can think of $t_\ell\propto a^\dagger_\ell$ and $t_\ell^\dagger\propto a$ as raising and lowering operators themselves. To go from the $b,c$ oscillators to the $a$ oscillators we are writing a partial Bogoliubov transformation (we have less $a$'s than $b$ and $c$ combined).

An important question is what is the normalization of the oscillators. This can be used to compute the expectation value of the number operators. This can be done by computing the commutator as follows
\begin{equation}
\vev{[t_\ell^\dagger, t_\ell]}_{\Omega} = \ell \sum_i r_i^{2\ell}- \ell  \sum_i \tilde r_i^{2\ell}=\ell(S_\ell-\tilde S_\ell)=\hbar_{eff}
\end{equation}
and on the right hand side we identify this with an effective $\hbar$ in the commutation relation. The expectation value of the number operator evaluated on the reference state $\ket \Omega$, which defines our vacuum,  is then
\begin{equation}
n_\ell=\vev{\hat N_{\ell}}_{\Omega} = \frac{\vev{t_\ell t_\ell^\dagger}_{\Omega}}{\hbar_{eff}}= \frac{\vev{a^\dagger a}_{\Omega}}{[a,a^\dagger]}=\frac{\tilde S_{\ell}}{S_\ell-\tilde S_\ell}
\end{equation}
Similarly we have that 
\begin{equation}
n_{\ell}+1=\vev{\hat N_{\ell}+1}_{\Omega}= \frac{S_{\ell}}{S_\ell-\tilde S_\ell}
\end{equation}

From these expectation values we can assign an entropy to the linear mode $a_\ell$. This is the entanglement entropy of the mode $a$ in the vacuum $\Omega$, and since the vacuum is obtained by a partial Bogoliubov transformation, we get that the reduced density matrix for the modes $a, a^\dagger$ look thermal. This entropy is given by
\begin{equation}
s_\ell = (n_{\ell}+1)\log(n_{\ell}+1)-n_{\ell} \log(n_\ell)
\end{equation}
which is again dependent on the geometric radii $r_i, \tilde r_i$. 

The meaning of this entropy is clear in the $N\to \infty$ limit where we are analyzing an effective field theory, but at finite $N$ it is more problematic because of all the cutoffs. One way to think about this entropy is that since the modes $a, a^\dagger$ act simply on the code subspace, they should induce an (approximate) factorization in the code subspace itself. More precisely, we can assign an entropy to an algebra  $\cal A$ acting on the reference state $\Omega$. The idea is that we need to produce a representation of the algebra by acting on $\Omega$ with the algebra, as follows ${\cal H_A}\simeq Span \{\CO \Omega \}$.

This induces a reduced density matrix for the ${\cal A}$ factor  such that 
\begin{equation}
\hbox{tr}(\rho {\cal O}) = \bra{\Omega} {\cal O}  \ket{\Omega}
\end{equation}
and we can associate the entropy $s_{\ell}$ with it.  If the algebra $\cal A$ acts in such a way that no element of the algebra annihilates the state, there is a second copy of ${\cal A}$, ${\cal A}^*$ that acts on ${\cal H_A}$ and commutes with $\cal A$. This second copy can be thought of as the thermal double of ${\cal A}$ (this is the Tomita-Takesaki theory, as discussed in \cite{Papadodimas:2013jku}). 
Roughly speaking, the state will look as follows
\begin{equation}
\sum_n \zeta_n \ket n \ket {\tilde n}
\end{equation}
where the $\zeta_n$ are a collection of  numbers and the $\ket n$ enumerate the possible states for the algebra $\cal A$ (for the Weyl algebra of a harmonic oscillator, there is a unique irreducible representation, so we enumerate the states by the occupation number). To avoid the infinite size representation and fit inside the code subspace,  we just need to truncate to states whose occupation number is bellow a cutoff induced by the code subspaces themselves.  The associated density matrix would be 
\begin{equation}
\rho \simeq \sum_n |\zeta_n|^2 \ket n \bra n
\end{equation}
and the dual algebra would act on the $\ket{\tilde n}$ states but not on the $\ket n$ states. For the states at hand, as long as the $\zeta_n\to 0$ sufficiently fast (which is usually true in an approximately thermal state), if the algebra is truncated or not becomes a moot point: the entropy is going to be dominated by the $\zeta_n$ where $n$ is small anyhow.

In the case above, we think of the representation space of ${\cal A}^*$ as a purification of $\rho$, so the ${\cal A}^*$ should be associated to the purification inside $\Omega$ for the generators of the oscillator algebra $a, a^\dagger$. Because we have a partial Bogoliubov transformation, we can compute this purification directly in terms of modes of the effective oscillators.

The idea is as follows.  Consider the two effective oscillators given by
\begin{eqnarray}
B_n^\dagger &=& \frac 1{\sqrt{S_n}}\sum r_i^n b_n^{(i)\dagger}\\
C_n^\dagger&=& \frac 1{\sqrt {\tilde S_n}} \sum \tilde r_i^n c_n^{(i)\dagger}
\end{eqnarray}
It is easy to check that these oscillators are normalized, and that the state $\ket\Omega$ is the ground state for the $B,C$ oscillators. From these oscillators it follows that 
\begin{equation}
t_n = \sqrt {S_n} B_n^\dagger -\sqrt{\tilde S_n} C_n
\end{equation}
and now it looks like part of a Bogoliubov transformation between only two modes. Normalizing the mode of the left hand side, which we will call $a^\dagger_n$, we need to take
\begin{equation}
a_n^\dagger := \frac {t_n}{\sqrt{\hbar_{eff}}} = \frac 1{\sqrt{S_n-\tilde S_n}} (\sqrt {S_n} B_n^\dagger -\sqrt{\tilde S_n} C_n)= \cosh (\gamma_n) B_n^\dagger-\sinh(\gamma_n) C_n
\end{equation}
and similarly for its adjoint.
The mode $a_n^\dagger$ is entangled with the mode
\begin{equation}
d_n^\dagger=\cosh(\gamma_n) C_n^\dagger-\sinh(\gamma_n)B_n
\end{equation}
that acts as the purification of the $a_n$ mode. Indeed, the $B,C$ oscillators can be recovered from $a,d$, and the ground state is a pure state of the $B,C$ modes. It is easy to check that the state $\ket \Omega$ in the $a,d$ basis is a squeezed state between the the $a,d$ modes. When tracing over $d$ we get a thermal density matrix for $a$, whose entropy is determined by the expectation value of the number operator: it maximizes the entropy given the constraint.

What is important for us is that the modes $B$, $C$ for moderately large $n$ are concentrated on the outermost edge of the concentric circle configuration, and the outermost anti-edge.
That is, most of the weight of the $B,C$ oscillators is concentrated on the modes at $r_1$ and $\tilde r_1$ respectively $b^{(1)},c^{(1)}$. The amplitudes for the other modes are exponentially suppressed in $n$. What this means is that for generic modes, the extrapolate dictionary can not penetrate beyond the geometric locus characterized by $\tilde  r_1$. By the Tomita -Takesaki analysis of \cite{Papadodimas:2013jku}, we can generate the states 
$\ket n _a\ket m_d$ from the state $\ket \Omega$ by acting with $a$ alone. In practice, this means we can recover the  algebra of the $d$ modes with an ensemble by having the reference ground state. 

Notice that the extrapolate dictionary seems to stop exactly at the outermost anti-edge. This was also suggested by the work of \cite{Balasubramanian:2017hgy}, which argued that there was an entanglement shadow in these geometries (a region where extremal Ryu-Takayanagi surfaces \cite{Ryu:2006bv} can not penetrate) and that the extremal surface that can enter the deepest stops exactly at this place.

In practice what this means is that the information inside the region of the outermost anti-edge is inaccessible. One can equally say that it is protected from quantum errors  generated by acting with the extrapolate dictionary. This information is encoded in  the modes $b,c$ that are orthogonal to $B,C$. 

Overall, the finite $N$ picture is similar to what we found before in \cite{Berenstein:2017abm, Berenstein:2016pcx}. The topology can be deduced from the uncertainties, although the procedure is more complicated. Also the entropy of the extrapolate dictionary modes is maximal given those uncertainties. Similar arguments can be used when we shift the reference ground state to a coherent state of the $b,c$ modes that fits comfortably inside the code subspace.

\section{Obstructions to having a globally well defined quantum metric}\label{sec:obs}

So far, we have found ourselves with a cutoff that is of order $\sqrt N$ on the modes, and we have assumed that the spacing between the radii is of order one.  From the point of view of Young tableaux, this is a situation where the lengths of the horizontal or vertical edges of the reference state are of order $N$. Our goal now is to push ourselves to a situation where we make some of these edges small enough so that the cutoff of $\sqrt N$ is already too large. That is, we want to take $M_i-M_{i+1}$ or $L_i-L_{i+1}$ to be of order $\sqrt N$ themselves. The idea now is to understand to what extent it is possible for us to define a metric operator in these setups. We only need to analyze the simplest case, with $L_1\simeq \sqrt N$, as in \eqref{eq:smallstep}, and we will call this  reference state $\ket {\Omega_1}$.
\begin{equation}
 \ytableausetup{boxsize=1.5 em}
\begin{ytableau}
\none[\ddots]&&& M_1\\
&&&\\
&&&L_1\\
\end{ytableau}\label{eq:smallstep}
\end{equation}
To analyze how one might get interference between the rows, we need to also analyze another reference state
\begin{equation}
 \ytableausetup{boxsize=1.5 em}
\begin{ytableau}
\none[\ddots]&&& S_1\\
&&&\\
&&&L_1\\
\end{ytableau}\label{eq:smallstep}
\end{equation}
with $S_1=M_1+1$. This reference state differs from the previous one by adding one column, and call this reference state $\ket{\Omega_2}$. The idea now is to look for a state that belongs to the code subspace generated from $\ket {\Omega_1}$ and $\ket {\Omega_2}$. The idea is to check if it is possible to find a globally defined metric that agrees between the two code subspaces as understood above in terms of building up perturbations relative to the reference state. The simplest such state is given by  
\begin{equation}
\begin{ytableau}
\none[\ddots]&&& M_1 &\\
&&&& \vdots\\
&&&&P\\
&&&L_1&\none[\vdots]\\
\end{ytableau}\label{eq:smallstep}
\end{equation}
where we have a column of length $P$ added to $\ket {\Omega_1}$ and that can be thought of as removing a column of length $L_1-P$ to $\ket{\Omega_2}$.

\begin{figure}[ht]
\includegraphics[width=15cm]{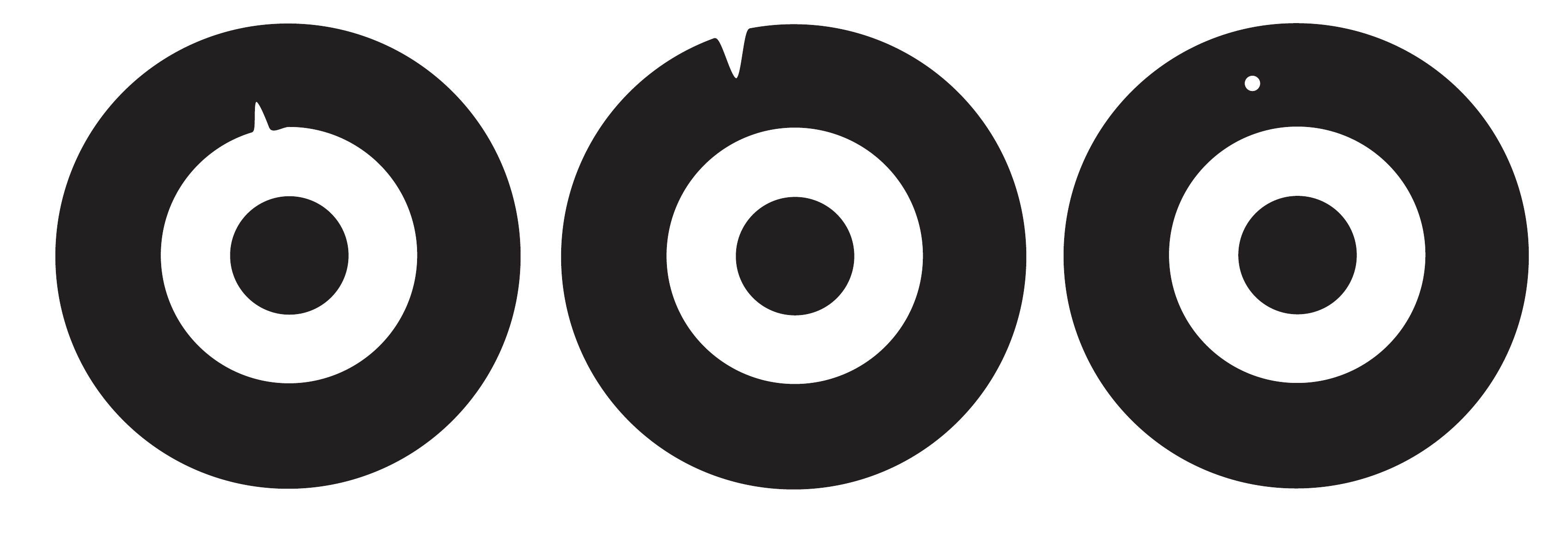}
\caption{LLM diagrams of three (identical) states with D-brane excitations that can be thought of as having nucleated in different ways in different code subsapces, and this results on them being pictured differently. }\label{fig:dbrane}
\end{figure}

The first state can be thought of as a superposition of states that have excitations around $\ket{\Omega_1}$ which are built from the $b^\dagger_{\Omega_1}$ modes. Such states are superpositions of excitations of
the top edge $r_1|_{\Omega_1}$. Relative to the reference state $\ket {\Omega_2}$, they are instead built by  superpositions of modes $c^\dagger_{\Omega_2}$, which originate in $\tilde r_1|_{\Omega_2}$. From the point of view of the two different code subspaces, we attach the excitation to different edges. This assignment is not local: we cannot specify the edge uniquely in a way that is independent of the reference state. The answer depends on the choice of reference state. We can even do this with superpositions of states of this type that lead to coherent states (for example, in \cite{Berenstein:2017abm} it was understood that a particular generating series of  these states is a coherent state).  Such generating series are of the type
\begin{eqnarray}
\sum_P \xi^P\ket P_{\Omega_1}\\
\sum_P \tilde \xi^{L_1-P} \ket{L_1-P}_{\Omega_2}
\end{eqnarray}
where $\xi$ is a complex number.
In order to get the same state, we need that $\xi = 1/\tilde \xi$. In our previous work, the range for $P$ was infinite, so convergence required that $|\xi|<1$ and similarly for $\tilde \xi$, so naively only one such state can be a nice coherent state. In practice, because of the cutoffs, the superposed coherent state does not belong completely to either of the two code subspaces nor is it  exactly a coherent state, but the state can have  a large overlap with states that do belong to either code subspace. The condition for large overlap is that $|\xi|<1$ or $|\tilde \xi|<1$. However, the state at fixed $P$ is a superposition of objects of either type.  Indeed, these objects are D-branes (giant gravitons \cite{McGreevy:2000cw}) that can be thought of as having nucleated at one edge (and belonging to it) and being moved to the other edge. This is shown in figure \ref{fig:dbrane}. Indeed, one can define a third code subspace that is a 
strip geometry plus a D-brane. In one code subspace the state is an excitation of $r_1$, in a second code subspace, the state is an excitation of $\tilde r_1$ and in the third code subspace the state is a strip geometry plus a D-brane (this is interpreted as a state with a different topology than the other two). 

Now, it is clear that there is no absolute boundary between these. The cutoffs are of order $\sqrt N$, but can be adjusted. What this means is that in practice $\ket {\Omega_1}$ and $\ket{\Omega_2}$ do not define a single code subspace. They define families of code subspaces that differ by the cutoffs. The state $\ket P_{\Omega_1}$ may or may not belong to either of these code subspaces.

Also notice that it gets more ambiguous when we try states of the form
\begin{equation}
\begin{ytableau}
\none[\ddots]&&& M_1 & &\\
&&&& \vdots & P_2\\
&&&&P&\none[\vdots]\\
&&&L_1&\none[\vdots]\\
\end{ytableau}\label{eq:smallstep}
\end{equation}
The first column is either an excitation of the $b$ modes or the $c$ modes depending on the code subspace, but the second column is an excitation of the $b$ modes in both code subspaces. 
The only condition on the Young tableaux is that $P_2\leq P$, but to belong to the different code subspaces, one needs to check that the cutoffs are not violated. We also find that
there is a new code subspace $\ket{\Omega_3}$, with $S_2=M_1+2$ where both of them would be assigned to excitations of $\tilde r_1|_{\Omega_3}$. We can keep on going this way so that the first $w_1$ columns are $c$ excitations and the next one is a $b$ excitation, versus all of them being $b$ excitations. one also gets code subspaces with one, two, up to $w_1$ D-branes if one wants to. 
This ambiguity makes it impossible to answer a question as to what is the metric as an operator that is independent of the choice of code subspace. Each code subspace has a different answer, and they are not compatible with each other.

In this case we can even say that the code subspaces with $D$-branes have different topology than the two original reference states. They also have a different spectrum of excitations: apart for the $b,c$ modes they also have the moduli of the D-branes themselves (these can be thought of as the additional coordinates $\xi= \tilde \xi^{-1}$ when we separate the D-branes from the code subspaces).

If one states that the metric information is encoded in the state as a message, what we are seeing is that different code subspaces that share the same state decode different messages. It is intriguing to speculate that different messages (different notions of the metric) are all allowed in the same sense that stringy dualities allow for more than one interpretation of the geometry, but only one of them will be sufficiently classical.  From what we have determined so far it is not yet clear that this is what is going on. So far we have done calculations in the absence of a concrete value for the string scale, relative to the Planck scale. To study the physics of the string scale would require studying modes that do not preserve as much supersymmetry. Such a problem is beyond the scope of the present paper.

One should be able to argue similarly for folded geometries.  These  geometries with folds are defined by stating that the number of $r,\tilde r $ variables is a function of the angle. The local supergravity analysis of the Poisson structure around each edge is the same \cite{Maoz:2005nk}, but one would have to define the mode expansions of the effective fields carefully. Different choices should be generically related by a linear transformation of the mode functions on each edge to a new set of functions. Such differences are accounted for by a Bogoliubov transformation
of the modes. There is no obvious preferred basis distinguished by the energy of the modes, because the reference state is no longer an eigenstate of the Hamiltonian. 
Unfortunately the dual field theory analysis is much more complicated because the construction of the corresponding dual states is not combinatorial. Our previous work  \cite{Berenstein:2016pcx, Berenstein:2017abm} dealt with these geometries in a particular approximation, but this approximation was not deduced from first principles. 
Studying this problem is very interesting as it should provide further details. Such analysis is beyond the scope of the present paper.

\section{Conclusion}

In this paper, we draw parallels between effective field theory, especially within the framework of the LLM geometries, and the notion of holographic code subspaces. We found that the nearby Hilbert space of states around some classical background, which is built by acting on the reference state with some number of effective fields results in a space defined in the same way as the code subspace developed in \cite{Almheiri:2014lwa}. It further matches the little Hilbert space of \cite{Papadodimas:2015jra}. We give explicit examples of code subspaces in the case of the LLM geometries, where we use concentric ring configurations as our reference state. To analyze this, we go beyond the infinite $N$ limit of our previous work. We show that the allowed effective fields are comprised of state dependent operators, insofar as analyzing the metric of a state depends on a choice of a code subspace in which to analyze it. 
 Further, we find that there is not a clear line between different code subspaces, and, in fact, there are states that clearly belong to multiple such  subspaces. This makes it ambiguous to write down a globally well defined metric operator, as the interpretation of how to obtain a metric   depends on the reference state that one builds the code space from. We have argued that this obstruction is essentially what forces us to state that it is only possible to interpret physics in an effective field theory of gravity within the framework of  code subspaces.
 If one thinks of the geometry of a quantum state as a quantum message, different code subspaces decode different messages from the same state. 

In this paper we dealt essentially exclusively with concentric ring LLM geometries. This is because the dual field theory states are well understood  as  combinatorial objects. In principle, a similar answer can be obtained for more general geometries, which can include folds (these are dealt in an approximate way in \cite{Berenstein:2017abm}). These  geometries with folds are characterized by the fact that the number of $r,\tilde r $ variables is a function of the angle. The supergravity analysis is more complicated because the modes necessarily mix in the extrapolate dictionary, and the cutoffs might depend non-trivially in the angle. These are interesting avenues of future research that can not be treated with the Young diagram technique. To address these, one should understand the cutoffs directly in the supergravity description.  

Another setup that is interesting is to deal with the superstar ensemble (as in \cite{Balasubramanian:2005mg}), which has properties more similar to a black hole. A big question here is to what extent we can use information the uncertainty and entropy in the extrapolate dictionary to make statements about geometry.
This is currently under research \cite{Weall}.

So far, all this work has been done for half BPS geometries. It would be very interesting to extend these ideas further to geometries that have less supersymmetry, or to excitations around such half BPS configurations that have less supersymmetry.  Such excitations could give additional insight into the more general structure of code subspaces and the corresponding cutoffs. Some of these can even be stringy states.

\acknowledgments

D.B. would like to thank V. Balasubramanian, J. de Boer, D. Kabat,  J. Maldacena, O. Parrikar, C. Rabideau for various discussions. 
Work  supported in part by the department of Energy under grant {DE-SC} 0011702.

\appendix

\section{Conjugacy classes of $S_n$ }\label{sec:appa}

 Consider the group of permutations $S_n$. The list of irreps of $S_n$ is in one to one correspondence with Young diagrams with $n$ boxes.
The cardinality of this set is also equal to the cardinality of the set of conjugacy classes of the group $S_n$, which we label by a group element representative $[\sigma]$.
The element $\sigma$ acts on the set of $n$ elements as a one to one function $\sigma: \{1, \dots n\}\to \{1, \dots, n\}$, sending $i \to \sigma(i)$. We can also represent this as a cycle decomposition
\begin{equation}
\sigma= (n^{(1)}_1 n^{(1)}_2\dots n^{(1)}_{k_1})(n^{(2)}_1\dots n^{(2)}_{k_2}) \dots
\end{equation} 
where the set of elements   $\{n^{(j)}_\ell \}$ is the set $\{1, \dots n\}$. It follows that $n= \sum k_i$
 Each $m\in \{1,\dots, n\}$ only appears once, and the elements on each parenthesis are called a cycle. 
We can recover the action on the set by the convention that $\sigma(n^{(j)}_{m})= n^{(j)}_{m+1}$, with $n^{(j)}_{k_j+1}\equiv n^{(j)}_{1}$. Basically, the cycles represent the iterated action of $\sigma$ on individual elements of the set $\{1, \dots n\}$. An individual cycle $ (n^{(1)}_1 n^{(1)}_2\dots n^{(1)}_{k_1})$ is said to have length $k_1$. We can choose the $k_i$ to be non-decreasing by permuting the order in which the individual cycles are presented. This does not change the assignment of $\sigma$.

\bibliography{LLMcodesubspaces}{}
\end{document}